\documentstyle[12pt,a4wide]{article}
\begin{document}
\thispagestyle{empty}
\begin{flushright}
DO-TH 95/07 \\
24 April 1995
\end{flushright}
\vspace{1cm}
\begin{center}
  \begin{Large}
{\bf INCLUSIVE SEMILEPTONIC DECAYS AND THE STRUCTURE OF B MESONS}
 \end{Large}
\end{center}
  \vspace{1cm}
   \begin{center}
C.\ H.\ Jin and E.\ A.\ Paschos\\
      \vspace{0.3cm}
        Institut f\"ur Physik,  Universit\"at Dortmund\\
        D--44221 Dortmund, Germany
  \vspace{1cm}
\end{center}
\renewcommand{\baselinestretch}{1.5}
\begin{abstract}
A field theoretic description for inclusive semileptonic B meson decays is
formulated. We argue that large regions of the phase spaces for the decays
are dominated by distances near the light cone. The light-cone dominance
allows to incorporate nonperturbative QCD effects in a distribution function.
A one-to-one correspondence with the heavy quark effective theory is
developed, which can estimate the first two moments of the distribution
function. These conditions are useful but not restrictive enough
to specify the distribution function,
which must still be determined from experiment. Several model-independent
predictions, such as scaling, sum rules of the hadronic structure
functions and relations among them, are made. General formulas for the
differential decay rates on several variables are presented, which are used
for calculating the electron energy spectra with an Ansatz for the light-cone
distribution function.
\end{abstract}
\newpage
\section{Introduction}
\renewcommand{\baselinestretch}{1.5}
\normalsize
The study of semileptonic and inclusive B meson decays is interesting
on several accounts. They are some of the simplest decays to study
theoretically and experimentally. For semileptonic processes
the structure of the lepton current is completely known and the inclusive
hadronic
tensor involves a sum over all final states, which makes the decay
products incoherent. Inclusive semileptonic B meson decays are
useful for obtaining parameters of the electroweak theory.
Two standard model parameters $V_{cb}$ and $V_{ub}$ are extracted from them.
In fact, the charmless inclusive semileptonic decay of B mesons is
the main experimental source at present which determines $V_{ub}$.
This follows from a measurement of the endpoint spectrum of the charged
lepton energy, whose better understanding is highly desirable.

The hadronic tensor for the decays involves short and long distance
contributions, where confinement effects are also important. There are,
however, several features of the decay which make its theoretical analysis
possible. The mass of the B-meson is large enough in comparison to 1 GeV,
so that the mass of the virtual W boson is large enough to produce a commutator
of two weak currents which are at light-cone distances relative to each other.
This occurs for large but not all regions of phase space, which allows us
to use the methods of deep inelastic scattering (DIS) and replace the
commutator of the two
currents with its singularity on the light cone times an operator bilocal
in the quark fields.
The matrix element of the bilocal operator between B-meson states contains
nonperturbative QCD corrections. Its light-cone Fourier transformation is
related to a distribution function, in direct analogy to DIS.
The distribution function
was discussed earlier \cite{bar,jin:pre,jin} and since it involves a heavy
quark it is expected to be large in the region where the b quark carries a
large fraction of the B-mesons momentum.

The hadronic matrix element has another feature which was
discussed recently. Namely, it contains a heavy quark and several properties
can be established in the heavy quark effective theory (HQET) \cite{hqet}.
In this paper we adopt the HQET and show that there is a correspondence
between results obtained by other groups \cite{chay,bigi,manohar,blok,mannel}
based on the HQET and moments of
the distribution function defined on the light cone. The first two moments
of the distribution function can be estimated, which determine the mean value
and the variance of it. These two properties are useful but not restrictive
enough to determine
the distribution function completely. We find that in order to reproduce the
decay spectrum, we need more detailed knowledge of the distribution function.
For this reason we find it appropriate to introduce the distribution
function over the whole range of its variable, instead of its first few
moments, and try to determine it from experiment.

In this paper we present a field theoretic prescription for inclusive
semileptonic B meson decays and determine properties of the distribution
function.
We intend to justify, in terms of field theory, the parton model for inclusive
B decays \cite{bar,jin:pre,jin} and to identify systematic procedures for
improving upon the parton model predictions. The advantage of this approach is
that model-independent aspects of the analysis can be clearly separated from
model-dependent ones. Several model-independent predictions are obtained.
Among them is scaling: away from the boundary of the phase space, the hadronic
structure functions depend on the scaling variable
\begin{displaymath}
\xi_+=[q\cdot P_B+\sqrt{(q\cdot P_B)^2-M_B^2(q^2-m_q^2)}]/M_B^2
\end{displaymath}
only,
and not on the momentum transfer squared $q^2$ directly.

The paper is planed as follows. In sect.~2 we give the kinematics and
the general formalism for inclusive semileptonic B meson decays. In sect.~3
we argue that very large domains of phase space are dominated by light-cone
distances of the two weak currents.
The distribution function is introduced in
sect.~4, while in sect.~5  the general properties and
the physical implication of the distribution function are discussed.
In sect.~6 we use the techniques of the operator product expansion (OPE) and
the HQET to estimate moments of the light-cone distribution function.
Sections 7 and 8 include predictions.
Some model-independent predictions are presented in sect.~7. We give the
triple differential decay rates for both $\bar B\rightarrow e\bar \nu_e X_u$
and $\bar B\rightarrow e\bar \nu_e X_c$ channels in sect.~8. In order to
produce
quantitative features of the data, we propose an Ansatz for the light-cone
distribution function consistent with its known properties. We evaluate the
electron energy spectra using the Ansatz. The conclusions are in sect.~9.

\section{Kinematics}
We consider the inclusive semileptonic decays $\bar B\rightarrow l\bar\nu_l
X_q$, where $l=e, \mu$, or $\tau$ and $X_q$ is any possible hadronic
final state containing a charm quark ($q=c$) or an up quark ($q=u$).
The decays
are produced by weak interactions. The partial decay width is given by
\begin{equation}
d\Gamma = \frac{1}{2E_B}\sum_{n}\left|{\cal M}\right|^2
\frac{d^3P_l}{(2\pi)^32E_l}\frac{d^3P_\nu}{(2\pi)^32E_\nu}
\Big[\prod^n_{i=1}\frac{d^3P_i}{(2\pi)^32E_i}\Big](2\pi)^4
\delta^4(P_B-q-\sum_{i=1}^nP_i) ,
\end{equation}
where $q$ is the four-momentum of the virtual W boson,
$P$ and $E$ denote the four-momentum and energy, respectively. Their
subscripts $B$, $l$, $\nu$, and $i$ correspond to the B meson, the
charged lepton, the neutrino, and the final state particle, respectively.
The summation on $n$  in Eq.(1) and hereafter implies
a sum over all possible final hadronic states
$\mid \mkern -6mu n \rangle$. At the tree level,
the decay amplitude is given by
\begin{equation}
{\cal M} = V_{qb} \frac{G_F}{\sqrt{2}}\bar u(P_l)\gamma^{\mu}
(1-\gamma_5)v(P_{\nu})
\langle n\left| j_\mu (0)\right| B\rangle ,
\end{equation}
where $V_{qb}$ are the matrix elements of the CKM matrix and the relevant
charged weak current is
\begin{equation}
j_\mu(x)=:\bar q(x)\gamma_\mu(1-\gamma_5)b(x): .
\end{equation}
$\mid \mkern -6mu B \rangle$ is the B-meson state with momentum $P_B^\mu$ and
is
normalized according to
$\langle B\mkern -6mu\mid \mkern -6mu B\rangle = 2E_B(2\pi)^3\delta^3({\bf
0})$.
Perturbative QCD corrections, which were studied
in \cite{ali,corbo,jez} and more recently in \cite{pert,kor},
are not included. They can be added in an analysis of the data as a
perturbation.
For unpolarized leptons
the partial decay width can be written as
\begin{equation}
d\Gamma= \frac{G_{F}^2\left|V_{qb}\right|^2}{(2\pi)^5E_{B}}L^{\mu\nu}
W_{\mu\nu}\frac{d^3P_{l}}{2E_{l}}\frac{d^3P_{\nu}}{2E_{\nu}} ,
\end{equation}
where $L_{\mu\nu}$ is the leptonic tensor
\begin{equation}
L^{\mu\nu}= 2(P^{\mu}_{l}P^{\nu}_{\nu}+P^{\mu}_{\nu}P^{\nu}_{l}-
g^{\mu\nu}P_{l}\cdot P_{\nu}+i\varepsilon^{\mu\nu}\hspace{0.06cm}_{\alpha\beta}
P^{\alpha}_{l}P^{\beta}_{\nu}).
\end{equation}
$W_{\mu\nu}$ is the hadronic tensor
\begin{equation}
W_{\mu\nu}=\sum_{n} \int\Big[\prod_{i=1}^n\frac{d^3P_i}{(2\pi)^3 2E_i}\Big]
(2\pi)^3\delta^4(P_B-q-\sum_{i=1}^n P_i)
\langle B\left| j_{\nu}^{\dagger} (0)\right| n\rangle \langle n\left|
j_\mu(0)\right|
B\rangle.
\end{equation}
It is useful to express the hadronic tensor in terms of a current commutator
\begin{equation}
W_{\mu\nu}= -\frac{1}{2\pi}\int d^4y e^{iq\cdot y}\langle B\left|[j_{\mu}(y),
j^{\dagger}_{\nu}(0)]\right| B\rangle .
\end{equation}
Thus the commutator of two weak currents is relevant in inclusive
semileptonic B meson decays. Furthermore, we shall see that the commutator
of currents $[j_\mu (y), j_\nu^\dagger (0)]$ near the light-cone
$y^2=0$ plays a central role in inclusive semileptonic B meson decays.

Generally, the hadronic tensor can be decomposed by
introducing the hadronic structure functions $W_{i}(q^2, q\cdot P_B)$ with
two scalar variables chosen to be $q^2$ and $q\cdot P_B$,
\begin{eqnarray}
W_{\mu\nu}(P_B, q) & = & -g_{\mu\nu}W_{1}(q^2, q\cdot P_B)
               +\frac{P_{B\mu}P_{B\nu}}{M_{B}^2}W_{2}(q^2, q\cdot P_B)
\nonumber\\
 &  & -i\varepsilon_{\mu\nu\alpha\beta}
\frac{P_{B}^{\alpha}q^{\beta}}{M_{B}^2}W_{3}(q^2, q\cdot P_B)+
    \frac{q_{\mu}q_{\nu}}{M_{B}^2}W_{4}(q^2, q\cdot P_B) \nonumber\\
 &  & +\frac{P_{B\mu}q_{\nu}+q_{\mu}P_{B\nu}}{M_{B}^2}W_{5}(q^2, q\cdot P_B) ,
\end{eqnarray}
where $M_B$ is the B-meson mass.
The hadronic structure functions $W_{i}(q^2, q\cdot P_B)$ characterize the
structure of the decaying B meson. Nonperturbative QCD effects for the
inclusive
process under consideration are incorporated in them.

Finally, the triple differential decay rate is obtained from the kinematical
analysis
\begin{equation}
\frac{d^3\Gamma}{dE_ldq^2dq_0}=\frac{G^2_F\left|V_{qb}\right|^2}{32\pi^3E_B}
L^{\mu\nu}W_{\mu\nu},
\end{equation}
where the contraction of the hadronic with the leptonic tensor yields
\begin{eqnarray}
L^{\mu\nu}W_{\mu\nu} & = & 2(q^2-M_l^2)W_{1}(q^2, q\cdot P_B) \nonumber\\
  &   & +[4P_l\cdot P_Bq\cdot P_B-4(P_l\cdot P_B)^2-M_{B}^2q^2+M_B^2M_l^2]
W_{2}(q^2, q\cdot P_B)/M_B^2 \nonumber\\
  &   & +2[(q^2+M_l^2)q\cdot P_B-2q^2P_l\cdot P_B]
W_{3}(q^2, q\cdot P_B)/M_B^2 \nonumber\\
  &   & +M_l^2(q^2-M_l^2)W_4(q^2, q\cdot P_B)/M_B^2 \nonumber\\
  &   & +4M_l^2(q\cdot P_B-P_l\cdot P_B)W_5(q^2, q\cdot P_B)/M_B^2,
\end{eqnarray}
where $M_l$ denotes the charged lepton mass.
There are three independent kinematical variables in this inclusive
phenomenology, for which we choose $P_l\cdot P_B$, $q\cdot P_B$,
and $q^2$.
We see that
for the massless lepton case only three hadronic structure functions
$W_1(q^2, q\cdot P_B), W_2(q^2, q\cdot P_B)$, and $W_3(q^2, q\cdot P_B)$
contribute. We will proceed to investigate the hadronic structure functions
in a way as suggested by the light-cone dominance.

\section{Light-Cone Dominance}
B mesons are heavy. This implies that its decay dynamics is analogous to
that of
deep inelastic scattering, that is, the light-cone dynamics dictates the
inclusive semileptonic B meson decay. To see this we start with an analysis
of the hadronic tensor.

According to the causality requirement the commutator in Eq.(7) has to vanish
for space-like $y$,
i.e.,
$[j_\mu(y),j_\nu^\dagger(0)]=0$,
for $y^2<0$, and hence the integrand in Eq.(7) has a support only for $y^2\geq
0$.

Taking $q=(q_0, 0, 0, q_3)$, we have
\begin{equation}
q\cdot y=q_0y_0-q_3y_3=\frac{1}{2}(q_0+q_3)(y_0-y_3)+\frac{1}{2}(q_0-q_3)
(y_0+y_3).
\end{equation}
The dominant contribution to the Fourier transform of the commutator in Eq.(7)
comes from domains with less
rapid oscillations, i.e. $q\cdot y={\cal O}(1)$; hence
\begin{equation}
y_0-y_3\sim \frac{1}{q_0+q_3},
\end{equation}
\begin{equation}
y_0+y_3\sim \frac{1}{q_0-q_3},
\end{equation}
and
\begin{equation}
y^2=y_0^2-y_1^2-y_2^2-y_3^2\leq y_0^2-y_3^2\sim \frac{1}{q_0^2-q_3^2}
=\frac{1}{q^2}.
\end{equation}
Therefore, the dominant contribution to the integral (7) results from the
range $0\leq y^2\leq 1/q^2$. This implies that as long as $q^2$ is large
enough, $q^2 \geq q_0^2$,
the decays take place
near the light-cone $y^2=0$, where $q_0^2$ is a reference scale and
experience shows that $q_0^2\simeq 1 \ GeV^2$. The scale $q_0^2$ should be
determined  ultimately by experiment.

For inclusive semileptonic B meson decays
$q^2$ varies in the physical range of
\begin{equation}
M_l^2\leq q^2\leq (M_B-M_{X_{min}})^2 ,
\end{equation}
where  $M_{X_{min}}$ is the minimum value of the final hadronic invariant mass.
The light-cone domain $q_0^2\leq q^2\leq (M_B-M_{X_{min}})^2$ covers most
of the phase space, since the B meson is so heavy that the interval
$(M_B-M_{X_{min}})^2-q_0^2\gg q_0^2-M_l^2$. Subsequently, in
inclusive semileptonic B meson decays the light-cone  contribution
dominates over all other nonperturbative QCD contributions. Contributions
far from the light cone are suppressed dynamically and kinematically. The
leading
approximation of nonperturbative QCD effects should be more reliable for the
charmless decays
$\bar B\rightarrow l\bar\nu_lX_u$, where $M_{X_{min}}$ is negligible, and/or
for
the decays to the final states containing a $\tau$ lepton with the mass
$M_\tau = 1.777 \ GeV$. For both cases
nonperturbative QCD contributions far from the light cone are  more
seriously suppressed kinematically.

\section{Distribution Functions}
Applying Wick's theorem one obtains
\begin{eqnarray}
[j_\mu(y), j_\nu^\dagger(x)] & = & \bar q(y)\gamma_\mu(1-\gamma_5)\{b(y),\bar
b(x)\}
\gamma_\nu(1-\gamma_5)q(x) \nonumber\\
&   & -\bar b(x)\gamma_\nu(1-\gamma_5)
\{q(x), \bar q(y)\}\gamma_\mu(1-\gamma_5)b(y) .
\end{eqnarray}
The quark pairs $q\bar q$ could be either charm or up quarks.
In the case of charm
quarks the matrix element between B mesons is very small, because there is
no spectator charm quarks in $B^-(b\bar u)$ and $\bar B^0(b\bar d)$ mesons.
The up-quark bilocal operator has a matrix element only between $B^-$ states,
but this term is suppressed for the reason given after Eq.(23).
For the matrix element we keep only the second term in Eq.(16).
The hadronic matrix element of the current commutator becomes
\begin{equation}
\langle B\left| [ j_{\mu}(y), j_{\nu}^{\dagger}(0) ]\right| B\rangle
= -\langle B\left|\bar b(0)
\gamma_{\nu} (1-\gamma_5)\{ q(0), \bar q(y)\}\gamma_{\mu} (1-\gamma_5)b(y)
\right
| B\rangle  .
\end{equation}
After some calculation, the expression (17) is transformed into
\begin{equation}
\langle B\left| [j_{\mu}(y), j_{\nu}^{\dagger}(0) ]\right| B\rangle
 = 2(S_{\mu\alpha\nu\beta}-
i\varepsilon_{\mu\alpha\nu\beta})[\partial^{\alpha} \Delta_q (y)]
\langle B\left| \bar b(0)\gamma^{\beta} (1-\gamma_5)b(y)\right
|B\rangle ,
\end{equation}
where $S_{\mu\alpha\nu\beta}=g_{\mu\alpha}g_{\nu\beta}+
g_{\mu\beta}g_{\nu\alpha}-g_{\mu\nu}g_{\alpha\beta}$. $\Delta_q(y)$ is the
Pauli-Jordan function for a free $q$-quark,
\begin{equation}
i\Delta_{q}(y)= \int \frac{d^4p}{(2\pi)^3}e^{-ip\cdot y}
  \varepsilon(p_{0})\delta (p^2-m_q^2) .
\end{equation}
In Eq.(18) the matrix element is separated in two factors. The first factor
contains the light-cone contribution in the form of the propagator and the
long-distance part is included in the reduced matrix element.
The decomposition is Lorentz covariant and each factor can be calculated
in the Lorentz frame of preference. This is analogous to deep inelastic
scattering where the production of two currents at light-like distances is
expanded in terms of operators times their Wilson coefficients, which are
obtained from perturbative QCD.

In the light-cone limit $y^2\rightarrow 0$,
the reduced matrix element in Eq.(18)
can be expanded in powers of $y^2$ from
the general arguments of Lorentz covariance and translation invariance:
\begin{equation}
\langle B\left|\bar b(0)\gamma^\beta (1-\gamma_5)b(y)\right|B\rangle
=4\pi P_B^\beta \sum_{n=0}^{\infty} (y^2)^nF_n(y\cdot P_B) .
\end{equation}
We define next the Fourier transform of
$F_n(y\cdot P_B)$,
\begin{equation}
\phi_n(\xi)=\int d(y\cdot P_B)e^{i\xi y\cdot P_B}F_n(y\cdot P_B) .
\end{equation}
Near the light cone only $\phi_0(\xi)$ survives,
defined as the Fourier transform of the reduced matrix element at
light-like separations
\begin{equation}
f(\xi)\equiv \phi_0(\xi)=\frac{1}{4\pi M_B^2}\int d(y\cdot P_B)e^{i\xi y\cdot
P_B}
\langle B\left| \bar b(0)/ \mkern -12mu P_B(1-\gamma_5)b(y)
\right|B\rangle \
|_{y^2=0}  .
\end{equation}
The physical implication and properties of this
distribution function will be discussed in the next section.

Finally, the leading contribution to the hadronic tensor
is obtained from Eqs.(7), (18) and (22):
\begin{equation}
W_{\mu\nu}=4(S_{\mu\alpha\nu\beta}-i\varepsilon_{\mu\alpha\nu\beta})
\int d\xi f(\xi) \varepsilon(\xi P_{B0}-q_0)\delta [(\xi P_B-q)^2-m_q^2]
(\xi P_B-q)^\alpha P_B^\beta  ,
\end{equation}
where $m_q$ is the mass of the quark in the final state.
Repeating the above steps for the first term in Eq.(16) we obtain
a distribution function for the up quark, whose $\xi$-variable is either
negative or very close to one. In the former case,
the negative $\xi$ is outside
the support of the distribution function and when $\xi$ is very close to
one the distribution function for a light-quark is very small, so that this
contribution will be neglected.

Comparing Eq.(23) with Eq.(8),
we are led to the expressions
\begin{eqnarray}
W_1(\xi_+, \xi_-) & = & 2[f(\xi_+)+f(\xi_-)] , \\
W_2(\xi_+, \xi_-) & = & \frac{8}{\xi_+ -\xi_-}[\xi_+f(\xi_+)-
                        \xi_-f(\xi_-)] , \\
W_3(\xi_+, \xi_-) & = & -\frac{4}{\xi_+ -\xi_-}[f(\xi_+)-f(\xi_-)] , \\
W_4(\xi_+, \xi_-) & = & 0 , \\
W_5(\xi_+, \xi_-) & = & W_3(\xi_+, \xi_-) ,
\end{eqnarray}
where the dimensionless variables $\xi_\pm$ are defined as
\begin{equation}
\xi_\pm = [q\cdot P_B\pm
\sqrt{(q\cdot P_B)^2-M_B^2(q^2-m_q^2)}]/M_B^2 .
\end{equation}
Hence, the light-cone dominance ascribes the hadronic structure functions to
a single universal light-cone distribution function.
The variable $\xi_-$ occurs for the first time in the decays of heavy particles
and is a consequence of field theory.

The triple differential decay rate (9) can then be written in terms of the
light-cone distribution function
\begin{eqnarray}
\frac{d^3\Gamma}{dE_ldq^2dq_0} & = &
\frac{G_F^2\left|V_{qb}\right|^2}{4\pi^3 E_B}
\frac{1}{\xi_+ -\xi_-} \Bigg \{ f(\xi_+)\Big[ (\xi_+ -\xi_-)
(q^2-M_l^2)/2 \nonumber\\
&   & +\xi_+[4P_l\cdot P_Bq\cdot P_B-4(P_l\cdot P_B)^2-M_B^2q^2+M_B^2
M_l^2]/M_B^2 \nonumber\\
&   & -[(q^2+3M_l^2)q\cdot P_B-2(q^2+M_l^2)P_l\cdot P_B]/M_B^2 \Big]
\nonumber\\
&   & -(\xi_+ \leftrightarrow \xi_-)\Bigg\} .
\end{eqnarray}

\section{Properties of the Light-Cone Distribution Function}
We discuss here some important properties of
the light-cone distribution function.
The distribution function
is normalized to unity:
\begin{eqnarray}
\int d\xi f(\xi) & = & \frac{1}{4\pi M_B^2}\int d\xi d(y\cdot P_B)
e^{i\xi y\cdot P_B}\langle B\left|\bar b(0)/ \mkern -12mu P_B(1-\gamma_5)b(y)
\right|B\rangle \mid_{y^2=0} \nonumber\\
                 & = & \frac{1}{2M_B^2}P_B^\mu\langle B\left|\bar b(0)
\gamma_\mu
(1-\gamma_5)b(0)\right| B\rangle = 1 ,
\end{eqnarray}
due to the conservation of the b  quantum number.

We consider next $f(\xi)$ in the rest frame of the B meson. In this frame,
\begin{equation}
f(\xi)=\frac{1}{2\pi}\int dy_0e^{i\xi M_By_0}
\langle B\left|b^\dagger (0)P_Lb(y_0)\right|B\rangle ,
\end{equation}
where the left-handed projection operator $P_L=(1-\gamma_5)/2$.
We insert a complete set of states between quark
fields, translate the $y_0$ dependence out of quark fields. Then we get
\begin{equation}
f(\xi)=\sum_m \delta
(M_B-\xi M_B-p_m^0) \left| \langle m\left| b_L(0)\right|
B\rangle \right|^2 ,
\end{equation}
where $b_L=P_Lb$. So we see that $f(\xi)$ obeys positivity. The state
$\mid \mkern -7mu m\rangle$ is
physical and must have $0\leq p_m^0\leq M_B$, thus $f(\xi)=0$,
for $\xi \leq 0$ and $\xi \geq 1$.
Therefore, the support of the light-cone distribution function reads
$0\leq \xi \leq 1$.
These
results are valid in any frame due to Lorentz invariance, although they are
deduced in the B rest frame. Furthermore, we observe from Eq.(33) that
$f(\xi)$ is the probability to find in the B meson a $b$ quark
with a momentum $\xi P_B$.
This is the familiar probabilistic interpretation of the
parton model, except it is written in the B rest frame rather than the
infinite momentum frame.

It will be convenient to expand the light-cone distribution function in terms
of derivatives of delta functions,
\begin{equation}
f(\xi)=\sum_{n=0}^{\infty}\frac{(-1)^n}{n!}M_n(\tilde{\xi})\delta^{(n)}
(\xi-\tilde{\xi}) .
\end{equation}
Such an expansion is very singular and any finite number of terms cannot
represent the differential decay width. The expansion is convenient for
comparisons with operator product expansions which also generate sequences
with singular terms.
The coefficient $M_n(\tilde{\xi})$ is related to
the nth moment about a point $\tilde{\xi}$ of the
distribution function as follows
\begin{equation}
M_n(\tilde{\xi})=\int d\xi  (\xi-\tilde{\xi})^nf(\xi) .
\end{equation}
It follows, now, that $M_0(\tilde{\xi})=1$, the mean value $\mu$
and the variance $\sigma^2$
of the light-cone distribution function can be expressed by the moments:
\begin{equation}
\mu\equiv M_1(0)=\tilde{\xi}+M_1(\tilde{\xi}) ,
\end{equation}
\begin{equation}
\sigma^2\equiv M_2(\mu)=M_2(\tilde{\xi})-M_1^2(\tilde{\xi}) .
\end{equation}

To sum up, our results so far are quite general.
It is shown that
the $b$-quark distribution function inside the B meson introduced in the
parton model \cite{bar,jin:pre,jin} is related to the light-cone Fourier
transformation
of the bilocal operator between B-meson states (22).
This makes clear the connection with the
parton model in inclusive B decays. In the next section we employ the
techniques of the operator product expansion and the heavy quark
effective theory to estimate moments of the light-cone distribution function.

\section{Moments of the Light-Cone Distribution Function}
To estimate moments of the light-cone distribution function,
we must calculate the hadronic matrix element
$\langle B\left| \bar b(0)\gamma^\beta (1-\gamma_5)b(y)\right|B\rangle$,
which involves long distances and hence brings in
nonperturbative effects of QCD. The techniques of
the OPE and the HQET provide a possibility to calculate it in a systematic
way.

We start with the light-cone OPE. Since the b quark is very heavy within the B
meson we can extract the large mass scale
\begin{equation}
b(y)=e^{-im_bv\cdot y} b_v(y) ,
\end{equation}
where $m_b$ is the b-quark mass and
$v$ is the velocity of the initial B meson, defined by
$P_B^\mu =M_Bv^\mu$ and the rescaled b-quark field $b_v(y)$ is related
to the effective field
of the HQET by Eq.(43) below.
Upon performing the light-cone OPE
the hadronic matrix element becomes
\begin{eqnarray}
\lefteqn{\langle B\left|\bar b(0)\gamma^\beta (1-\gamma_5)b(y)\right| B\rangle
=} \nonumber\\
& &\mkern -25mu e^{-im_bv\cdot y}\sum_{n=0}^\infty \frac{(-i)^n}{n!}
y_{\mu_1}\cdots y_{\mu_n}
\langle B\left|\bar b_v(0)\gamma^\beta (1-\gamma_5)
{\cal S}[k^{\mu_1}\cdots k^{\mu_n}]
b_v(0)\right|B\rangle
\end{eqnarray}
with $k_\mu =iD_\mu$.
${\cal S}$ denotes the symmetrization.
This OPE keeps the leading twist operators with
higher twist effects being neglected.
The advantage of this expansion is twofold. First, the Lorentz structure of
the matrix element allows us to express it in terms of the B-meson momentum
\begin{eqnarray}
\lefteqn{\langle B\left|\bar b_v(0)\gamma^\beta (1-\gamma_5)
{\cal S}[k^{\mu_1}\cdots k^{\mu_n}]
b_v(0)\right| B\rangle =} \nonumber\\
 & & 2(C_{n0}P_B^\beta P_B^{\mu_1}\cdots P_B^{\mu_n}+
\sum_{i=1}^nM_B^2C_{ni}g^{\beta\mu_i}P_B^{\mu_1}\cdots P_B^{\mu_{i-1}}
P_B^{\mu_{i+1}}\cdots P_B^{\mu_n})+ \nonumber\\
& & terms \ with \ g^{\mu_i\mu_j} .
\end{eqnarray}
Terms with $g^{\mu_i\mu_j}$ can be omitted on the light cone.
Second, we can estimate some terms in the HQET as shown below.
Substituting Eqs.(39) and (40) into Eq.(22) we have
\begin{equation}
f(\xi)=\sum_{n=0}^\infty \frac{(-1)^n}{n!}(\sum_{i=0}^n C_{ni})
\delta^{(n)}(\xi-\frac{m_b}{M_B}) .
\end{equation}
Comparing with Eq.(34), the nth moment about the point $\tilde\xi =m_b/M_B$ of
the light-cone distribution function is related to the expansion coefficients
as following
\begin{equation}
M_n(m_b/M_B)=\sum_{i=0}^n C_{ni} .
\end{equation}
It is straightforward to note that the moment
$M_0(m_b/M_B)=C_{00}$ is exactly equal to 1.

We now go to the second step further. We employ the HQET
to estimate other expansion
coefficients.
In this effective theory the rescaled b-quark field $b_v(x)$ is expressed
by the
velocity-dependent heavy quark field $h_v(x)$ by means of an expansion in
powers of $1/m_b$,
\begin{equation}
b_v(x)=[1+\frac{i/ \mkern -12mu D}{2m_b}+
{\cal O}(\frac{1}{m_b^{2}})]h_v(x) ,
\end{equation}
where $D_\mu$ is the covariant derivative.
The effective Lagrangian is
\begin{equation}
{\cal L}_{HQET}= \bar h_viv\cdot Dh_v+\bar h_v\frac{(iD)^2}{2m_b}
h_v
 -\bar h_v\frac{gG_{\alpha\beta}\sigma^{\alpha\beta}}{4m_b}
h_v+{\cal O}(\frac{1}{m_b^2}) ,
\end{equation}
where $igG_{\mu\nu}=[D_\mu, D_\nu ]$.
The series expansion in powers of $1/m_b$ is now explicit.

By virtue of the methods based on the
HQET the expansion coefficients $C_{ni}$ in Eq.(40)
can be expressed in terms of
small quantities, proportional to powers $\Lambda_{QCD}/m_b$;
to be precise, the order of $C_{ni}$
and hence that of the moment $M_n(m_b/M_B)$ is
expected to be $(\Lambda_{QCD}/m_b)^n$. Hence nonperturbative effects can,
in principle, be calculated in a systematic manner.
A few coefficients are calculated to be
\begin{eqnarray}
C_{10} & = & \frac{5m_b}{3M_B}E_b +{\cal O}(\Lambda_{QCD}^3/m_b^3) , \\
C_{11} & = & -\frac{2m_b}{3M_B}E_b+{\cal O}(\Lambda_{QCD}^3/m_b^3) , \\
C_{20} & = & \frac{2m_b^2}{3M_B^2}K_b+{\cal O}(\Lambda_{QCD}^3/m_b^3) , \\
C_{21} & = & C_{22} \ = \  0 ,
\end{eqnarray}
where the dimensionless parameters, which parametrize  the nonperturbative
phenomena, are defined as \cite{manohar}
\begin{eqnarray}
K_b & \equiv & -\frac{1}{2M_B}\langle B\left|\bar h_v
\frac{(iD)^2}{2m_b^2}h_v
\right|B\rangle , \\
G_b & \equiv & \frac{1}{2M_B}\langle B\left|\bar h_v\frac{gG_{\alpha\beta}
\sigma^{\alpha\beta}}{4m_b^2}h_v\right| B\rangle ,
\end{eqnarray}
with $E_b=K_b+G_b$.
Both parameters are expected to be order
$(\Lambda_{QCD}/m_b)^2$.

According to Eqs.(45-48) and (42), the first two moments of the light-cone
distribution function
are
\begin{eqnarray}
M_0(m_b/M_B) & = & 1 , \\
M_1(m_b/M_B) & = & \frac{m_b}{M_B}E_b+{\cal O}(\Lambda_{QCD}^3/m_b^3) , \\
M_2(m_b/M_B) & = & \frac{2m_b^2}{3M_B^2}K_b+{\cal O}(\Lambda_{QCD}^3/m_b^3) .
\end{eqnarray}
$M_0$ is exactly equal to 1.
$M_1(m_b/M_B)$ receives no contribution of order $\Lambda_{QCD}/m_b$.
As a consequence, there are no nonperturbative QCD corrections to
moments at the level $\Lambda_{QCD}/m_b$; they arise
first at order $(\Lambda_{QCD}/m_b)^2$.

Substituting Eqs.(52) and (53) into Eqs.(36) and (37),
the mean $\mu$ and the variance $\sigma^2$ of the light-cone
distribution function $f(\xi)$ are estimated up to $(\Lambda_{QCD}/m_b)^2$
corrections to be
\begin{equation}
\mu = \frac{m_b}{M_B}(1+E_b) ,
\end{equation}
\begin{equation}
\sigma^2 = (\frac{m_b}{M_B})^2(\frac{2K_b}{3}-E_b^2).
\end{equation}
Therefore, the light-cone distribution function $f(\xi)$ is sharply peaked
around $\xi =\mu \approx m_b/M_B$ and its width is of order
$\Lambda_{QCD}/M_B$, in agreement with rather general expectations.

We can furthermore implement a numerical analysis.
The parameter $G_b$ for the B meson can be related to observables
\cite{manohar}
\begin{equation}
m_bG_b= -\frac{3}{4}(M_{B^\ast}-M_B).
\end{equation}
The experiment determines $G_b$ to be $-0.0065$. The parameter $K_b$
was estimated using a QCD sum rule \cite{ball}.
It has a large uncertainty and its range could be $K_b=0.006-0.012$.
In Tables 1 and 2 we list $\mu$ and $\sigma^2$ for various parameters
$m_b$ and $K_b$, respectively.
The numerical evaluation indicates that $\mu =0.85-0.95$
and $\sigma^2 =0.003-0.007$, should the b-quark mass and the parameter $K_b$
vary within the limits $4.5 \ GeV\leq m_b\leq 5.0 \ GeV$ and
$0.006\leq K_b\leq 0.012$.

A few remarks are in order:

(1) The free-quark decay model is reproduced, if one keeps only the first
term in the series (41), i.e., $f(\xi)=\delta (\xi-m_b/M_B)$.
The nonperturbative corrections manifest themselves in the second and higher
terms in Eq.(41).

(2) Important information about the light-cone distribution function is
obtained by
applying the OPE and the HQET. The first two moments of it are reduced to
two accesible parameters $G_b$ and $K_b$.
However, it should be emphasized that the
first few moments do not exhaust the information hidden in the
distribution function,
because they do not determine it completely. This point
becomes obvious when one observes
that any truncated resummation of the expansion in Eq.(34) cannot produce
a smooth function.
This is the origin of the singularity at the upper endpoint of the electron
energy spectrum found in \cite{bigi,manohar,blok},
where a truncated HQET-based OPE has been
used.
Moreover, a truncated series gives
rise to a delta function in the hadronic tensor $W_{\mu\nu}$,
which demands the
decay to be described by quark kinematics instead of hadron
kinematics. This brings about ambiguities particularly at the endpoint of
the $b\rightarrow u$ electron energy spectrum.
In quark kinematics the
endpoint of the $b\rightarrow u$ electron energy spectrum
lies at $E_e=m_b/2$, while the actual endpoint should be
$E_e=M_B/2$ from kinematics at hadron level.

As a matter of fact, an infinite number of terms in the light-cone OPE (39)
must be included, and we cannot reduce our task to the calculation of a few
matrix elements of lower dimension operators.
It should be noted that, when more moments are taken into
account, then higher dimensional operators of the OPE are involved;
although  qualitatively the
moment $M_n(m_b/M_B)$ is expected to be of order $(\Lambda_{QCD}/m_b)^n$
in the framework of the OPE and the HQET,
their hadronic matrix elements are much more difficult
to calculate.
We may conclude that the OPE and the HQET can serve
as an useful technique for
obtaining additional information on nonperturbative QCD, but they alone are
not sufficient to determine the shape of the heavy quark light-cone
distribution function. The limitations of this method must be complemented
by other theoretical approaches.
Alternatively, the required information may become available from experiment.
For example, the light-cone distribution function can be
determined from a measurement of the triple differential decay rate
$d^3\Gamma/dE_edq^2d\xi_+$, as advocated in \cite{moriond}.

(3) A resummation of the operator product expansion has been performed
recently in \cite{neubert,resum,mn} in order to eliminate the
difficulties mentioned previously in the approach
for inclusive B decays \cite{bigi,manohar,blok}.
Their treatments using
an operator product expansion in the context of the $1/m_b$ expansion
is different in an essential way
from what we have formulated here. The distribution function is introduced
in section 4 in a general way (without invoking the HQET),
in analogy to DIS.
In the present work the HQET is employed to estimate moments of
the light-cone distribution function. To this end one may use other
nonperturbative approaches (e.g. QCD sum rules). Moreover, the
physical interpretation of distribution functions is distinct: the structure
function $f(k_+)$ defined in \cite{mn} determines the probability to find a $b$
quark with the light-cone residual momentum $k_+$ inside the B meson, while the
light-cone distribution function $f(\xi)$ defined by Eq.(22) is the probability
of finding
a $b$ quark carrying momentum $\xi P_B$ within the B meson.
The differences originate  drastically different predictions.

\section{Model-independent Predictions}
The light-cone distribution function contains the long-distance physics
associated with strong interactions of the $b$ quark inside the B meson.
Although we know some important properties of this function derived on
general grounds and remarkable progress was made in calculating its
first few moments,
it cannot yet be determined completely from
first principles.
Nevertheless, it is interesting to draw model-independent
results without relying on the quantitative aspect of distribution functions.

It is convenient to define the scaling structure functions
$F_i$
\begin{eqnarray}
F_1(\xi_+,q^2) & = & \frac{1}{2}W_1(\xi_+,q^2) ,\\
F_2(\xi_+,q^2) & = & \frac{\xi_+ -\xi_-}{8}W_2(\xi_+,q^2) ,\\
F_3(\xi_+,q^2) & = & \frac{\xi_+ -\xi_-}{4}W_3(\xi_+,q^2) ,\\
F_5(\xi_+,q^2) & = & \frac{\xi_+ -\xi_-}{4}W_5(\xi_+,q^2) .
\end{eqnarray}
The kinematical analysis shows that in the light-cone domain
and away from the resonance region, namely in the kinematical region
where our approach applies, $f(\xi_-)$ is expected to be relatively
small and can be ignored, since the light-cone distribution function is
sharply peaked around $\xi =\mu \approx m_b/M_B$, as established by the HQET
analysis in the last section.
We anticipated this, because the function $f(\xi_-)$ describes the
creation of a quark-antiquark pair inside the B meson
through Z-diagram and the virtual correction should be small.
Then Eqs.(24-28) are simplified to
\begin{equation}
\xi_+ F_1(\xi_+,q^2) =  F_2(\xi_+,q^2) = - \xi_+ F_3(\xi_+,q^2)
= - \xi_+ F_5(\xi_+,q^2)
= \xi_+ f(\xi_+) .
\end{equation}
Two important features of these expressions are:\\
(i) the structure functions $F_i$ satisfy scaling: they become functions of
$\xi_+ = [q\cdot P_B + \sqrt{(q\cdot P_B)^2 -M_B^2(q^2-m_q^2)}]/M_B^2$
alone and are independent of the momentum transfer squared $q^2$;\\
(ii) the structure functions are related to each other through the
light-cone distribution function.

Thus the structure functions $F_i(\xi_+,q^2)$ are measures of the momentum
distribution
of the b quark in the decaying B meson.
The first result is the analogue of the Bjorken scaling in B-decays.
The second one will be evidence
for the spin-$1/2$ nature of charged partons (the quarks), i.e. the analogue
of the Callan-Gross relation.
Furthermore, using Eq.(61) and the normalization of the light-cone
distribution function leads to the following sum rules:
\begin{eqnarray}
\lefteqn{\int_0^1 d\xi_+ F_1(\xi_+,q^2)= \int_0^1 \frac{d\xi_+}{\xi_+}
F_2(\xi_+,q^2) = -\int_0^1 d\xi_+ F_3(\xi_+,q^2)} \nonumber\\
& & = -\int_0^1 d\xi_+ F_5(\xi_+,q^2) = \int_0^1 d\xi_+ f(\xi_+) = 1 .
\end{eqnarray}
These results are very similar to those in DIS, since the
behavior of the hadronic tensors for both cases is dictated by the light-cone
dynamics.
The remarkable thing is that these results follow in general without having
information about the specific shape of the light-cone distribution
function. It is important to keep in mind that the above results are valid
up to perturbative and non-leading nonperturbative QCD corrections.
As in DIS including the perturbative QCD corrections will lead to scaling
violation: quantities which scale will be modified by powers of $lnq^2$.
In order to uncover these properties a detailed measurement of the
differential decay rate $d^3\Gamma/dE_ldq^2dq_0$, Eq.(9), and hence of the
structure functions $F_i(\xi_+,q^2)$ is essential.

\section{Electron Energy Spectra}
The results obtained so far take into account the lepton mass effects.
Now we concentrate on the inclusive semileptonic B decay to electrons,
in which the electron mass is negligible. The differential
decay rates for $\bar B\rightarrow e\bar \nu_eX_q$ in the B rest frame
is then simplified from Eq.(30) to be
\begin{equation}
\frac{d^3\Gamma}{dE_edq^2dq_0}=\frac{G_F^2\left|V_{qb}\right|^2}{4\pi^3}
\frac{q_0-E_e}{\sqrt{{\bf q^2}+m_q^2}} \{f(\xi_+)(2E_e\xi_+-q^2/M_B)-
(\xi_+ \leftrightarrow \xi_-)\} ,
\end{equation}
where the variables $\xi_{\pm}$ given in (29) become
\begin{equation}
\xi_{\pm}=\frac{q_0\pm \sqrt{{\bf q^2}+m_q^2}}{M_B}
\end{equation}
with $q= c, u$ relevant to the $b\rightarrow c$ and the
$b\rightarrow u$ decays,
respectively.

As already pointed out, appropriately including nonperturbative QCD effects
allows us to use the actual kinematical limits at hadron level and to give
the correct $E_e$ end point. They are given by:
\begin{eqnarray}
0\leq & E_e & \leq \frac{M_B}{2}(1-\frac{M^2_{X_{min}}}{M_B^2}) ,\\
0\leq & q^2 & \leq 2E_e(M_B-\frac{M^2_{X_{min}}}{M_B-2E_e}) ,\\
E_e+\frac{q^2}{4E_e}\leq & q_0 & \leq \frac{q^2+M_B^2-M^2_{X_{min}}}{2M_B} .
\end{eqnarray}
In Figs.~1 and 2 the phase spaces for the $b\rightarrow c$ and
the $b\rightarrow u$
decays are demonstrated, respectively, which show also domains of validity for
our approach. In the resonance region bound-state effects in the final state
become large. However, physical quantities, integrated over an appropriate
phase space region, could be calculated reliably in our approach.

In order to calculate decay distributions, one needs an Ansatz for the
light-cone distribution function $f(\xi)$ consistent with the general
properties pointed out in section 5. We propose a parametrization for the
light-cone distribution function with two parameters $a$ and $b$ as follows
\begin{equation}
f(\xi)=N\frac{\xi (1-\xi)}{(\xi-b)^2+a^2} \ ,
\end{equation}
where N is the normalization constant. In addition, constraints on the
parameters $a$ and $b$ are imposed by the numerical evaluation of
the mean value and the variance
of the light-cone distribution function
implemented in section 6 based on
the techniques of the OPE and the HQET. Within the bounds of Tables 1 and 2,
we find $a=0.002-0.016$ and $b=0.86-0.97$.
Our light-cone distribution function is illustrated in Fig.~3.

Next, we use the distribution function (68) to compute
the electron energy spectra for both $b\rightarrow c$ and $b\rightarrow u$
decays, shown in Figs.~4 and 5 respectively. We see that both spectra are
smooth and go to zero at the endpoint. There is no sharp spike as $E_e$
approaches the upper endpoint.
A desirable consequence is that perturbative QCD corrections
on top of our spectra will give finite results, i.e.~without
endpoint singularities, because of the vanishing of the spectra at the
endpoint. Therefore
the endpoint behaviour of the spectra with perturbative QCD corrections
are smooth for both
$b\rightarrow u$ as well as $b\rightarrow c$ decays.

Comparing with the free-quark decay model, we find that nonperturbative
QCD corrections appear to be significant for the spectra in the
endpoint region, referring to Figs.~4 and 5.
In particular, nonperturbative QCD effects become
more important for the charmless $\bar B\rightarrow e\bar \nu_eX_u$
decay spectrum
in the endpoint region. Our $b\rightarrow u$ endpoint spectrum is considerably
softer than the free-quark decay spectrum, which is finite and non-zero
at the endpoint.

\section{Conclusions}
We studied the semileptonic and inclusive B meson decays using field theoretic
methods, which justifies several steps of the parton model.
We give in terms of a light-cone distribution function the general formulas
for the $\bar B\rightarrow l\bar \nu_l X_q$
decays, keeping the mass of the final quark $m_q=m_u$ or $m_c$.
These formulas should be useful for analysing the decay spectra and
determining the CKM matrix elements $\left|V_{cb}\right|$ and
$\left|V_{ub}\right|$.

Additional properties of the distribution function are derived from the
HQET. These are the first two moments. They are useful but do not determine
the distribution function completely. The distribution function must be
determined, presently, from detailed fits of the data. For this reason we
discuss a new parametrization of the distribution function, which
in the heavy quark limit reduces
to a delta function and thus reproduces the free-quark decay model.

There is now at our disposal a complete and reliable formalism for inclusive
semileptonic B meson decays. It can be used directly to fit experimental data.
This could be done, preferably, by experiment groups, which can appropriately
include their experimental conditions.
We are also working along the line trying to reproduce the electron energy
spectrum and other features of the data.
Early comparisons of the parton-model spectrum with experimental data are
encouraging \cite{jin}.

The approach is more reliable for the charmless $\bar B\rightarrow e
\bar \nu_e X_u$ decays and shows that a large percentage of the events
(more than $15\%$) has $E_e>2.3$ GeV, which may result only from the
$b\rightarrow u$ transition. It should be possible to analyse the
endpoint spectrum in our approach including radiative corrections and
obtain a reliable value for $V_{ub}$.

\bigskip
\bigskip
\begin{flushleft}
{\bf Acknowledgements}
\end{flushleft}
We wish to thank W. Palmer who participated at the early stages of this work
and C.S. Huang for useful discussions on the HQET. One of us (EAP) thanks
A. Vainshtein for helpful discussions emphasizing the close connection
between the light-cone and the HQET approaches.
The support of BMBF is gratefully acknowledged (05-6DO93P).

\newpage
Table 1 The mean $\mu$ of the light-cone distribution function for
several values of $m_b$ and $K_b$.
\begin{center}
\begin{tabular}[t]{lcccccc} \hline \hline
$m_b[GeV]$ & 4.5 & 4.6 & 4.7 & 4.8 & 4.9 & 5.0\\ \hline
$K_b=0.006$ & 0.849 & 0.867 & 0.886 & 0.905 & 0.924 & 0.943\\
$K_b=0.007$ & 0.849 & 0.868 & 0.887 & 0.906 & 0.925 & 0.944\\
$K_b=0.008$ & 0.850 & 0.869 & 0.888 & 0.907 & 0.926 & 0.945\\
$K_b=0.009$ & 0.851 & 0.870 & 0.889 & 0.908 & 0.927 & 0.946\\
$K_b=0.010$ & 0.852 & 0.871 & 0.890 & 0.909 & 0.928 & 0.947\\
$K_b=0.011$ & 0.853 & 0.872 & 0.891 & 0.910 & 0.929 & 0.948\\
$K_b=0.012$ & 0.854 & 0.873 & 0.892 & 0.911 & 0.930 & 0.948\\ \hline \hline
\end{tabular}
\end{center}
\vspace{1cm}
Table 2 The variance $\sigma^2$ of the light-cone distribution
function for several values of $m_b$ and $K_b$.
\begin{center}
\begin{tabular}[t]{lcccccc} \hline \hline
$m_b[GeV]$ & 4.5 & 4.6 & 4.7 & 4.8 & 4.9 & 5.0\\ \hline
$K_b=0.006$ & 0.00288 & 0.00301 & 0.00314 & 0.00328 & 0.00342 & 0.00356\\
$K_b=0.007$ & 0.00336 & 0.00351 & 0.00367 & 0.00383 & 0.00399 & 0.00415\\
$K_b=0.008$ & 0.00384 & 0.00401 & 0.00419 & 0.00437 & 0.00456 & 0.00474\\
$K_b=0.009$ & 0.00432 & 0.00451 & 0.00471 & 0.00492 & 0.00512 & 0.00533\\
$K_b=0.010$ & 0.00480 & 0.00501 & 0.00523 & 0.00546 & 0.00569 & 0.00592\\
$K_b=0.011$ & 0.00527 & 0.00551 & 0.00575 & 0.00600 & 0.00625 & 0.00651\\
$K_b=0.012$ & 0.00574 & 0.00600 & 0.00627 & 0.00654 & 0.00681 & 0.00709\\
\hline \hline
\end{tabular}
\end{center}
\newpage
\begin{center}
{\bf Figure Captions}
\end{center}
\vspace{1cm}
\noindent
1. The phase space for the $b\rightarrow c$ inclusive semileptonic
decay of B-mesons. The dashed and dotted lines encircles
the domain of validity for our approach.
The region between the solid and the dashed curves is the resonance
region. \hfill\\
2. Same as Fig.1, but for the $b\rightarrow u$ decay.\\
3. The light-cone distribution function (68). The parameters are taken to
be $a=0.0076$ and $b=0.92$.\\
4. The electron energy spectrum in  $\bar B\rightarrow e\bar \nu_eX_c$
decays. The solid line is obtained in our approach for $M_{X_{min}}=m_c=
1.5 \ GeV$, $a=0.0076$ and $b=0.92$. The dashed line corresponds to the
free-quark decay model spectrum, using $m_b=5.0 \ GeV$ and $m_c=1.7 \ GeV$
set by a fit to ARGUS data.\\
5. Same as Fig.4, but for the $b\rightarrow u$ decay. The solid line is
predicted in our approach for $M_{X_{min}}=m_u=0$, $a=0.0076$ and $b=0.92$.
The dashed line results from the free-quark decay model, using
$m_b=5.0 \ GeV$ and $m_u=0$.
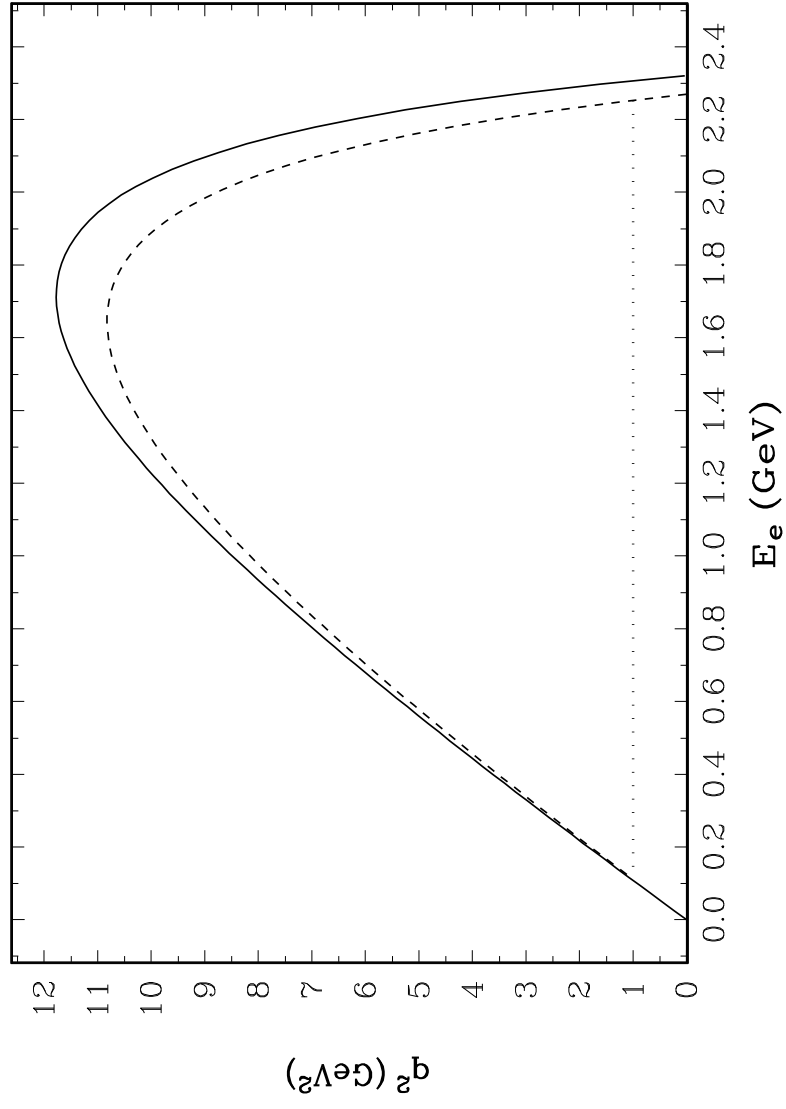
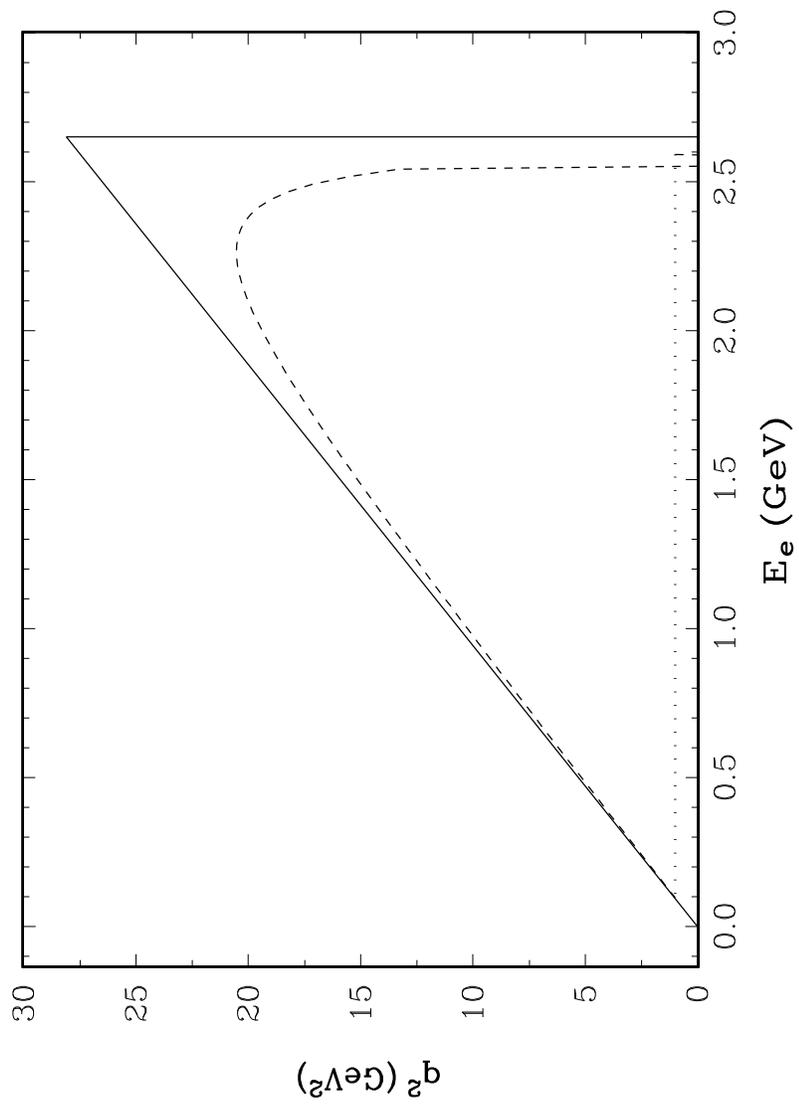
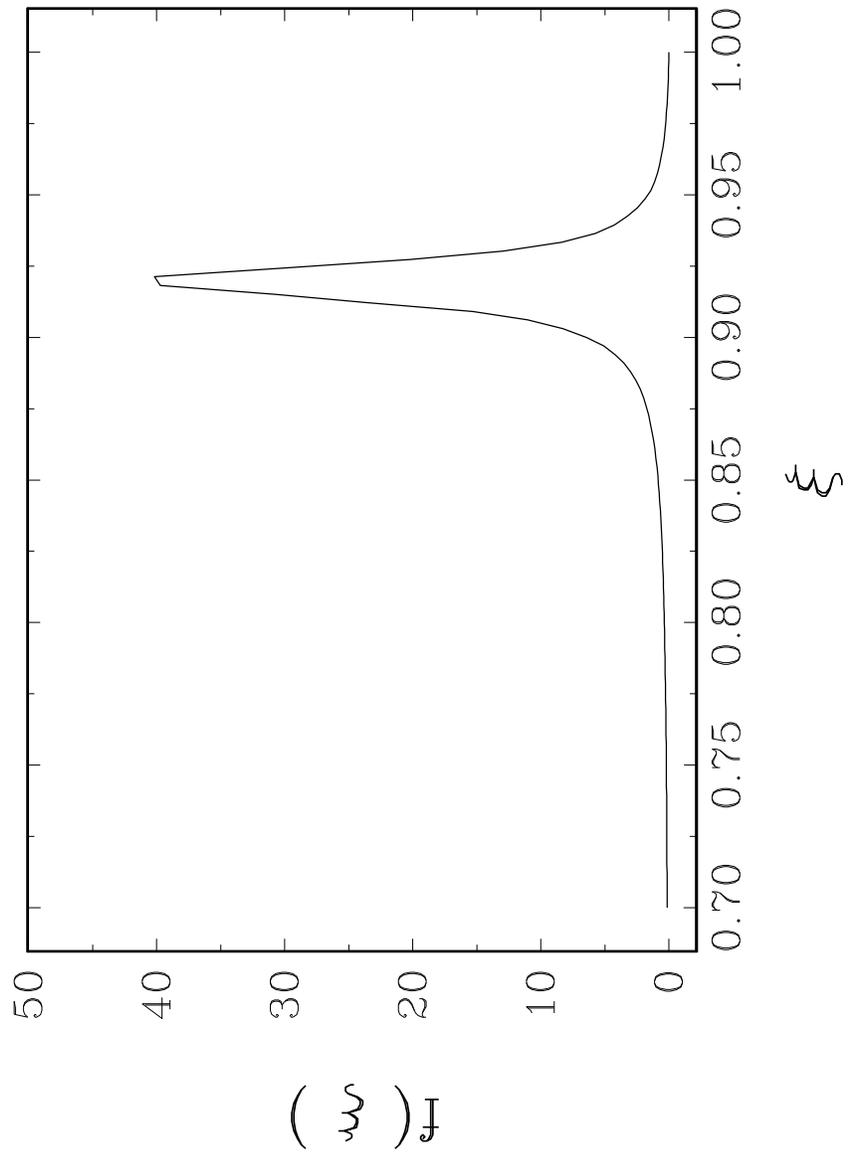
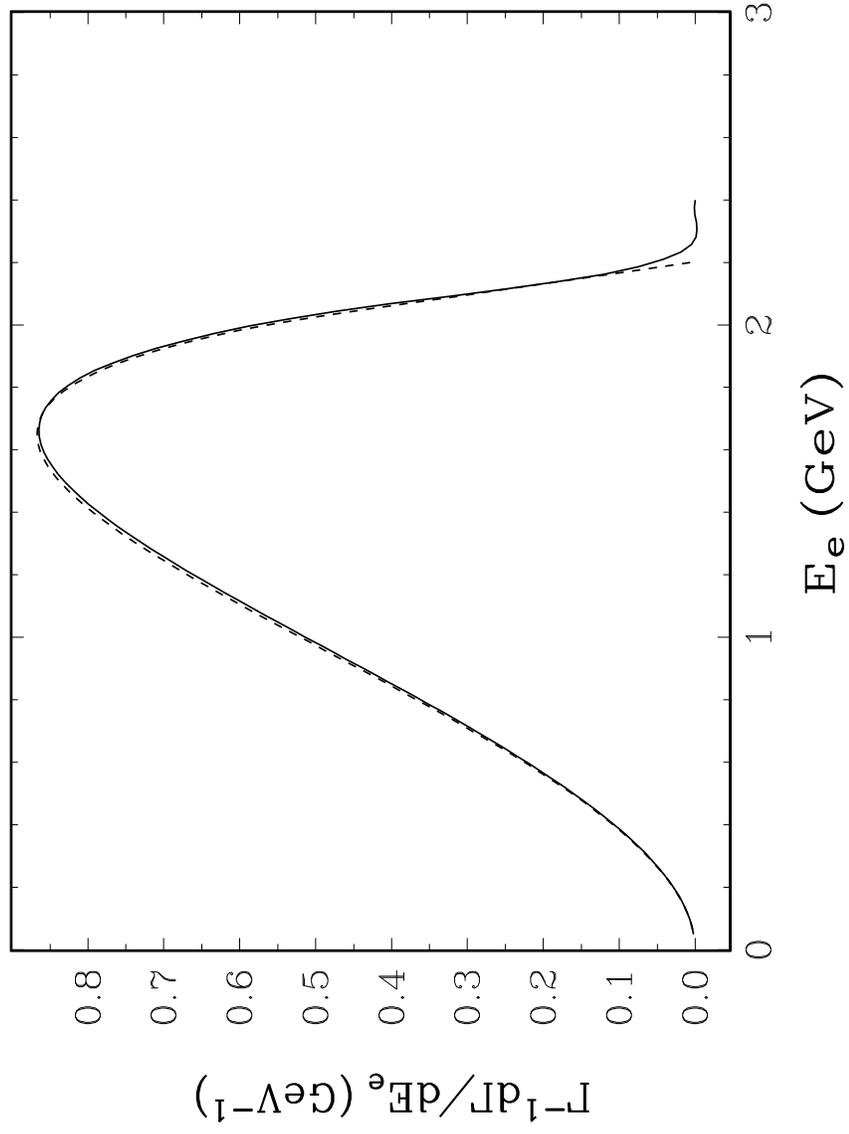
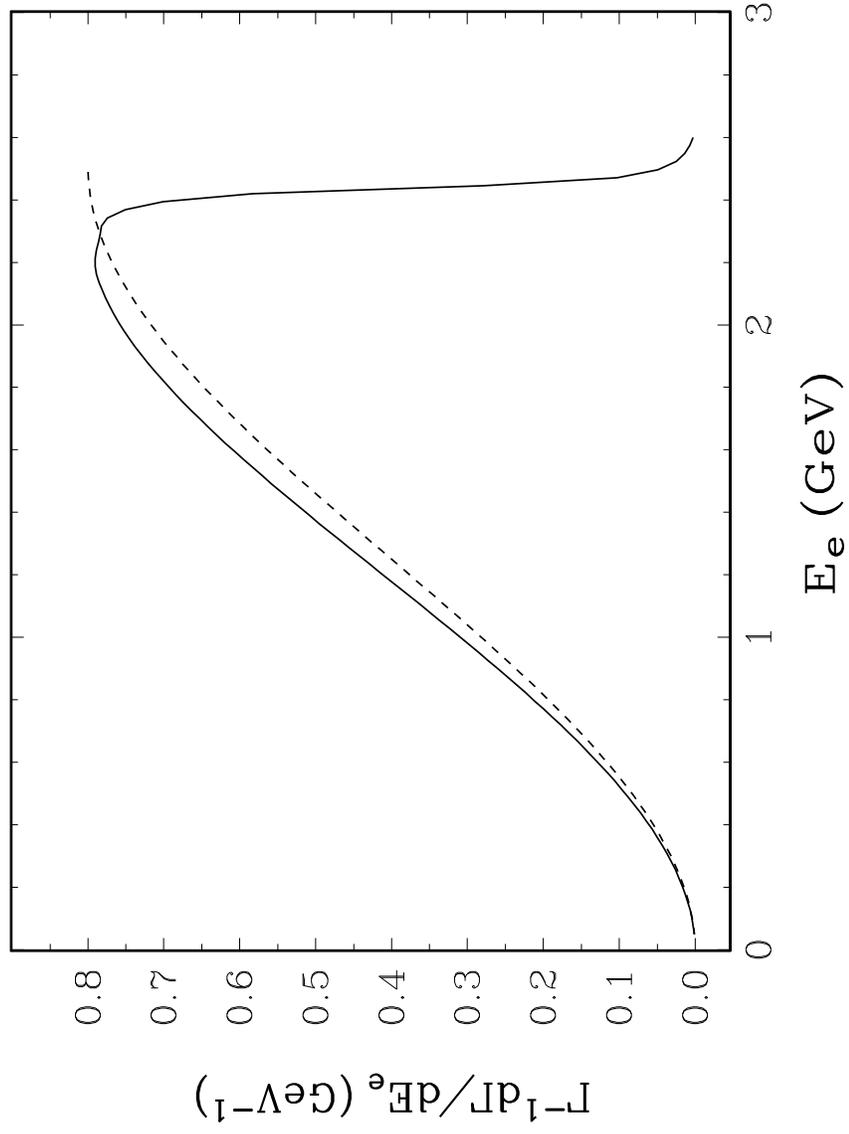
\end{document}